\documentclass[10pt,journal,compsoc]{IEEEtran}
\ifCLASSOPTIONcompsoc
  \usepackage[nocompress]{cite}
\else
  \usepackage{cite}
\fi
\ifCLASSINFOpdf
   \usepackage[pdftex]{graphicx}
\else
\fi
\usepackage{amsmath}
\usepackage{algorithm}
\usepackage{algorithmic}
\usepackage{amsmath}  

\usepackage{multirow}

\begin{document}
%
% paper title
% Titles are generally capitalized except for words such as a, an, and, as,
% at, but, by, for, in, nor, of, on, or, the, to and up, which are usually
% not capitalized unless they are the first or last word of the title.
% Linebreaks \\ can be used within to get better formatting as desired.
% Do not put math or special symbols in the title.
\title{Evaluation Mappings of Spatial Accelerator Based On Data Placement}
%
%
% author names and IEEE memberships
% note positions of commas and nonbreaking spaces ( ~ ) LaTeX will not break
% a structure at a ~ so this keeps an author's name from being broken across
% two lines.
% use \thanks{} to gain access to the first footnote area
% a separate \thanks must be used for each paragraph as LaTeX2e's \thanks
% was not built to handle multiple paragraphs
%
%
%\IEEEcompsocitemizethanks is a special \thanks that produces the bulleted
% lists the Computer Society journals use for "first footnote" author
% affiliations. Use \IEEEcompsocthanksitem which works much like \item
% for each affiliation group. When not in compsoc mode,
% \IEEEcompsocitemizethanks becomes like \thanks and
% \IEEEcompsocthanksitem becomes a line break with idention. This
% facilitates dual compilation, although admittedly the differences in the
% desired content of \author between the different types of papers makes a
% one-size-fits-all approach a daunting prospect. For instance, compsoc 
% journal papers have the author affiliations above the "Manuscript
% received ..."  text while in non-compsoc journals this is reversed. Sigh.

% \author{Michael~Shell,~\IEEEmembership{Member,~IEEE,}
%         John~Doe,~\IEEEmembership{Fellow,~OSA,}
%         and~Jane~Doe,~\IEEEmembership{Life~Fellow,~IEEE}% <-this % stops a space
\author{Wu Zhipeng,
        Liu Yu,~\IEEEmembership{Member,~IEEE,}% <-this % stops a space
\IEEEcompsocitemizethanks{\IEEEcompsocthanksitem Liu Yu was with the School of Microelectronics, Tianjin University, Tianjin,
China, 300072.\protect\\
% note need leading \protect in front of \\ to get a newline within \thanks as
% \\ is fragile and will error, could use \hfil\break instead.
E-mail: liuyu@tju.edu.cn
\IEEEcompsocthanksitem Wu Zhipeng with Tianjin University.}% <-this % stops an unwanted space
% \thanks{Manuscript received April 19, 2005; revised August 26, 2015.}
}

\IEEEtitleabstractindextext{%
\begin{abstract}
The scheduling strategies of workloads are critical to fully exploiting the performance of spatial accelerators, accurate performance models are required to evaluate the mapping of workloads.
Recent works proposed various cost-model to describe the dataflow of the spatial accelerator. 
However, they are less expressive about customized memory hierarchies and thus lead to inaccurate performance models. 
In this paper, we propose, PolyAcc, a framework for evaluating the mappings of workload on spatial accelerator based on data placement. 
The Data placement relation describes the temporal-spatial relation of data at different memory levels, which can accurately capture the runtime behavior of hardware units. 
Based on data placement relations, polyAcc accurately analyzes the data volume for different reuse patterns and estimate metrics, including data reuse, latency, and energy. 
Overall, polyAcc closely matches the ideal execution time and PE utilization for GEMM and Conv workloads, respectively achieves 0.82\%, 18.8\% improvements for execution time and energy consumption estimates in validation against Eyeriss architecture compared to the state-of-the-art framework.

\end{abstract}

% Note that keywords are not normally used for peerreview papers.
\begin{IEEEkeywords}
Spatial Accelerator, Cost-Model, Polyhedral Model
\end{IEEEkeywords}}

% make the title area
\maketitle

% To allow for easy dual compilation without having to reenter the
% abstract/keywords data, the \IEEEtitleabstractindextext text will
% not be used in maketitle, but will appear (i.e., to be "transported")
% here as \IEEEdisplaynontitleabstractindextext when the compsoc 
% or transmag modes are not selected <OR> if conference mode is selected 
% - because all conference papers position the abstract like regular
% papers do.
\IEEEdisplaynontitleabstractindextext
% \IEEEdisplaynontitleabstractindextext has no effect when using
% compsoc or transmag under a non-conference mode.

% For peer review papers, you can put extra information on the cover
% page as needed:
% \ifCLASSOPTIONpeerreview
% \begin{center} \bfseries EDICS Category: 3-BBND \end{center}
% \fi
%
% For peerreview papers, this IEEEtran command inserts a page break and
% creates the second title. It will be ignored for other modes.
\IEEEpeerreviewmaketitle

\IEEEraisesectionheading{\section{Introduction}\label{sec:introduction}}
% Computer Society journal (but not conference!) papers do something unusual
% with the very first section heading (almost always called "Introduction").
% They place it ABOVE the main text! IEEEtran.cls does not automatically do
% this for you, but you can achieve this effect with the provided
% \IEEEraisesectionheading{} command. Note the need to keep any \label that
% is to refer to the section immediately after \section in the above as
% \IEEEraisesectionheading puts \section within a raised box.

% The very first letter is a 2 line initial drop letter followed
% by the rest of the first word in caps (small caps for compsoc).
% 
% form to use if the first word consists of a single letter:
% \IEEEPARstart{A}{demo} file is ....
% 
% form to use if you need the single drop letter followed by
% normal text (unknown if ever used by the IEEE):
% \IEEEPARstart{A}{}demo file is ....
% 
% Some journals put the first two words in caps:
% \IEEEPARstart{T}{his demo} file is ....
% 
% Here we have the typical use of a "T" for an initial drop letter
% and "HIS" in caps to complete the first word.
% \IEEEPARstart{T}{his} demo file is intended to serve as a ``starter file''
% for IEEE Computer Society journal papers produced under \LaTeX\ using
% IEEEtran.cls version 1.8b and later.
% % You must have at least 2 lines in the paragraph with the drop letter
% % (should never be an issue)
% I wish you the best of success.

% \hfill mds
 
% \hfill August 26, 2015
% Tensor operation have been widely deployed in many application, such as data analysis, deep learning, which are often compute- and memory-intensive.
Spatial accelerators have emerged in the recent past to accelerate compute- and memory-intensive applications, such as data analysis, deep learning, due to their lower runtime and energy-efficiency. 
% The typical spatial accelerator is a hierarchical architecture. As shown in Fig.\ref{fig:example}(a), spatial accelerator is parallelism using numerous processing elements~(PEs), efficiency communication using on-chip network to connect those PEs, and aggressive data reuse using private/shared scratchpad with efficient scheduling.
The main factor for the high efficiency of the spatial accelerators is that allocate dedicated memory to form customized memory hierarchies that ensure the compute engine is busy computing the data being fed, namely, the "right" data should be moved in or out the "right" locations at the "right" time-stamp.
Programmers must manage critical data orchestration explicitly to precisely control when and where data is used. 
A primary current focus is how to estimate performance of mappings of workload on spatial accelerators.

To evaluate the performance of accelerators, Some recent works have been proposed various cost models. State-of-the-art techniques  estimate performance and energy consumption using either compute-centric~\cite{timeloop,Interstellar,dMazeRunner} or data-centric~\cite{maestro}, or relation-centric~\cite{tenet}. However, both framework have some limitations.
First, these analytical methods are less expressive, and they can only represent a subset of the topology of the accelerator, incomplete abstractions could hurt the accurate performance model. For example, failure to consider interconnections between hardward units limits data reuse analysis of compute-centric method. And, the previous work~\cite{maestro} assume a fixed accelerator architecture with a 2-level memory hierarchy, which limits the framework's use.
Second, both analytical methods fail to support accurate performance analysis. The compute-centric does not consider data transfer between PEs, and the data-centric and relation-centric ignore the impact of memory, leading to inaccurate performance estimates.

All of these limitations affect the quality of evaluating results and further impact the mapping decisions that designers derive based on these results. To overcome this challenge, in this paper, we propose a framework, PolyAcc, to estimate the performance and energy consumption of mappings on spatial architectures. We present a memory-centric representation to model the spatial accelerator, which is a hierarchical tree to describe various memory hierarchies and interconnections in the accelerator.
We propose data placement relation~(DPR) to capture the runtime behavior of data transfer between hardware units across space and time.
Based on the DPR, performance metrics can be easily computed using polyhedral operations. Overall, polyAcc can estimate various hardware metrics, including data reuse, latency, PE utilization, and energy consumption. Results show that PolyAcc achieves accurate PE utilization estimates, 0.82\% and 18.8\% improvements in execution time and energy consumption estimates compared to the state-of-the-art framework. 
Overall, the contribution of the paper are:
\begin{itemize}
    \item A memory-centric representation to model the spatial accelerator and mapping strategies. 
    \item An analytical model for mapping workload on a spatial architecture with data placement relation.
    \item A closed-form analytical method to estimate workload execution time and energy. 
\end{itemize}

The rest of the paper is organized as follows. Section 2 introduces the background of spatial accelerator and polyhedral model. Section 3 presents the PolyAcc framework. Section 4 presents experimental results. Finally, this paper is concluded in Section 5.

\section{Background}
\subsection{Spatial Accelerator}

\begin{figure}[h]
	\centerline{\includegraphics[width=1\linewidth]{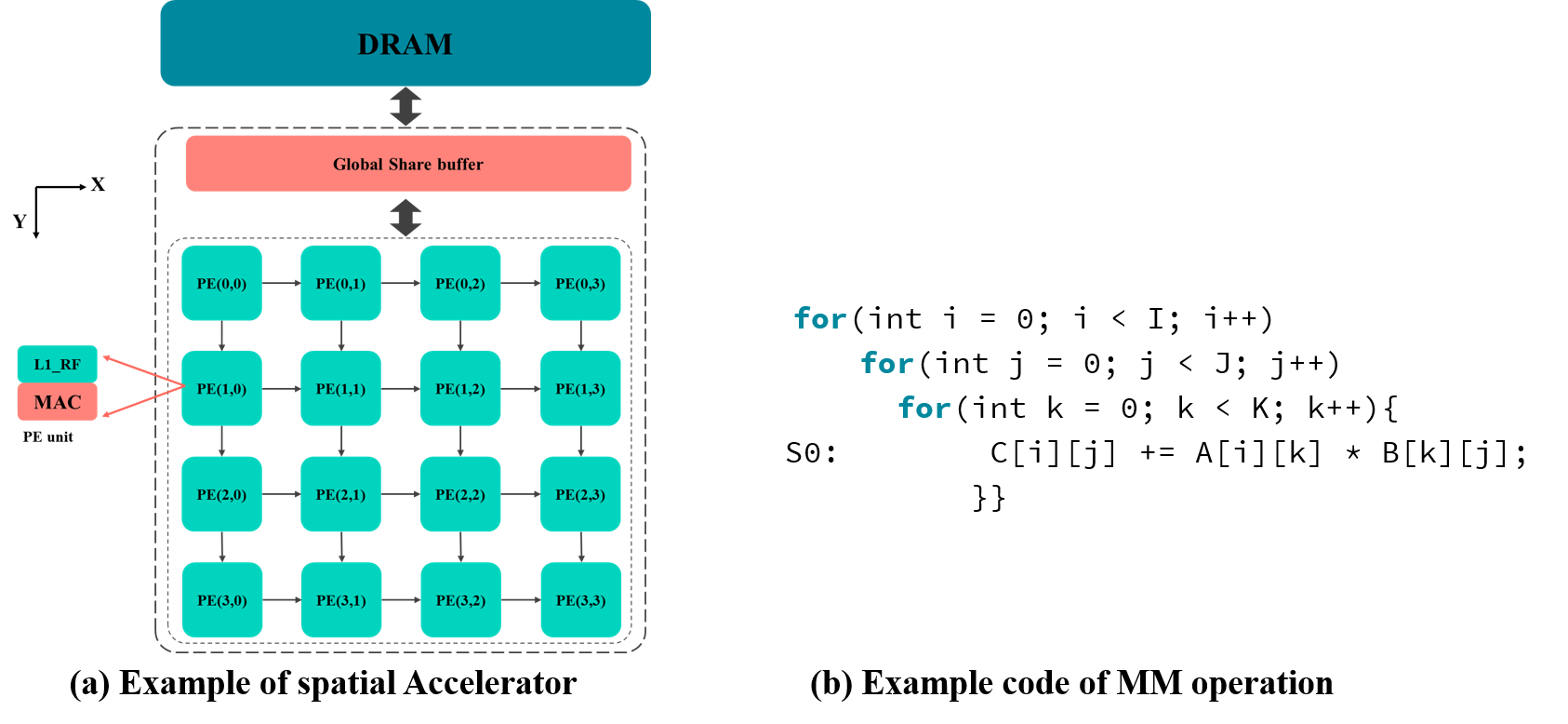}}
	\caption{Example of the Spatial Accelerator and MM operation. (a)shows an example of a spatial architecture with output-stationery dataflow, which has three levels of memory: DRAM, on-chip shared memory and private register files in PEs. (b)show the example code of MM operator}
	\label{fig:example}
\end{figure}

The spatial accelerator is mainly composed of a set of processing elements(PE) and the interconnection between PEs. 
Each PE includes local scratchpad and arithmetic logic units (ALUs). 
Fig.\ref{fig:example}(a) shows the example of a spatial accelerator with output-stationery dataflow, which is composed of $4\times4$ PE array, a on-chip shared memory, and private register files for each PE. Each PE is connected to the PE to the left and below, which represent data communication between PEs.

% Programmers must explicitly orchestrate data, controlling when and where hardware units read and write data, data can be reused across different memory levels to reduce memory access. 
Based on the complex memory hierarchy and connection relation of hardware unit in the spatial accelerators, there are three major patterns of data reuse: 
\emph{(1)Temporal Reuse}: 
units access the same data set stored in themselves across time. 
\emph{(2)Spatial Reuse}: 
multiple units access the same data set at the same time via multicast/broadcast connection 
\emph{(3)Spatial-Temporal Reuse}: 
adjacent units with point-to-point connections access the same data set in a skewed schedule.

These reuse features must be considered when evaluating the mapping strategies for any algorithm running on spatial accelerators. 

\subsection{Polyhedral Model}
The polyhedral model is a powerful abstraction for analyzing and transforming loop nest programs. 
A program in the polyhedral model is typically represented by three items: iteration domain, access function, and a schedule. The matrix multiplication code shown in Fig.\ref{fig:example}(b) will be used to illustrate these concepts.

The iteration domain contains the loop instances of the statement in the program. The iteration domain of the loop in Fig.\ref{fig:example}(b) is defined as the following: $domain=\{S0[i, j, k] : 0\leq i\leq I \land 0\leq j\leq J \land 0\leq k\leq K \}$, where $S0[i, j, k]$ is a loop instance and $0\leq i\leq I \land 0\leq j\leq J \land 0\leq k\leq K$ gives the constraints.

The access function map the statement instance to the array index, which include read access function and write access function. The read access function in Fig.\ref{fig:example}(b) are shown below: $R=\{S0[i, j, k]\to C[i, j]; S0[i, j, k]\to B[k, j]; S0[i, j, k]\to A[i, k]\}$. The write access relation in the loop is: $W=\{S0[i, j, k]\to C[i, j]\}$.
The access function means the loop instance $S0[i, j, k]$ access the array element $ C[i, j]$, $B[k, j]$ and $A[i, k]$.

The schedule represents the lexicographical order of statement instances that leverages a multi-dimensional affine function. The schedule in Fig.\ref{fig:example}(b) are shown below: $Sch=\{S0[i, j, k]\to [i, j, k]\}$.
The schedule can be explicitly encoded using a tree structure\cite{schedule_tree}, which can simplify the modeling of automatic memory managements in polyhedral compilers.

In this work, we use the Integer Set Library\cite{isl} for performing polyhedral operations, and we use the same notation as used in ISL to elucidate the concepts and the algorithm. Table~\ref{tab:notation} list the notation for polyhedral operations used in the article.

\begin{table}[h]
    \centering
    \footnotesize
    \renewcommand{\arraystretch}{1.5}
    \setlength{\tabcolsep}{3pt}
    \caption{The notation for Polyhedral operations used in the article }
    \label{tab:polyhedral_notation}
    \scalebox{.8}{
        \begin{tabular}{cc}
        \hline
        \multicolumn{1}{c|}{Notation}   & Interpretation               \\ \hline
        \multicolumn{1}{c|}{$\circ$}  & Two relations are composed to form a new relation.                           \\
        \multicolumn{1}{c|}{${R}^{-1}$} & The relation $R$ is inversed.    \\
        \multicolumn{1}{c|}{$\cup$}     & The union of  two integer sets or relations.        \\
        \multicolumn{1}{c|}{$\cap$}     & The intersection of two integer sets or relations. \\
        \multicolumn{1}{c|}{$Sum()$} &  Building a counting formula for an integer set or relation. 
        \\ \hline
        \end{tabular}
        }
\end{table}

\section{Overview of PolyAcc}
\begin{figure}[h]
	\centerline{\includegraphics[width=\linewidth]{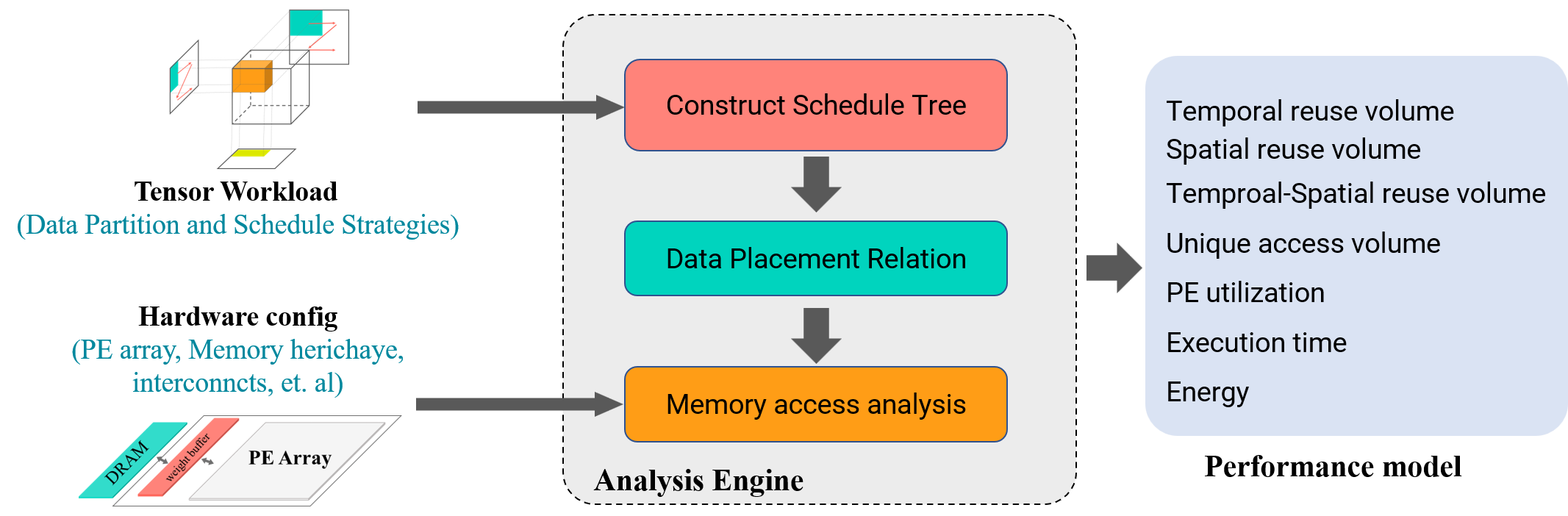}}
	\caption{PolyAcc Overview}
	\label{fig:overview}
\end{figure}

In this section, we describe our framework, PolyAcc, for implementing mapping analysis for the spatial accelerator based on data placement.
As shown in Fig.\ref{fig:overview}, PolyAcc takes the specification of spatial architecture, the data partition and schedule strategies as input.
Then, PolyAcc automatically constructs a scheduling tree including memory hierarchy information, which represent the lexicographical order of the
workload.
After that, PolyAcc scans the scheduling tree to infer each memory level's data placement relation. PolyAcc analyzes data reuse via data placement relations and obtains various performance metrics precisely, which can help designers explore the hardware design space and find the optimal mapping strategies.

% Analysis engine computes the data access information in each level memory, performance model computes various performance metrics based on the reuse information.

\subsection{Abstraction for Architecture and Mapping}
\begin{figure}[h]
	\centerline{\includegraphics[width=\linewidth]{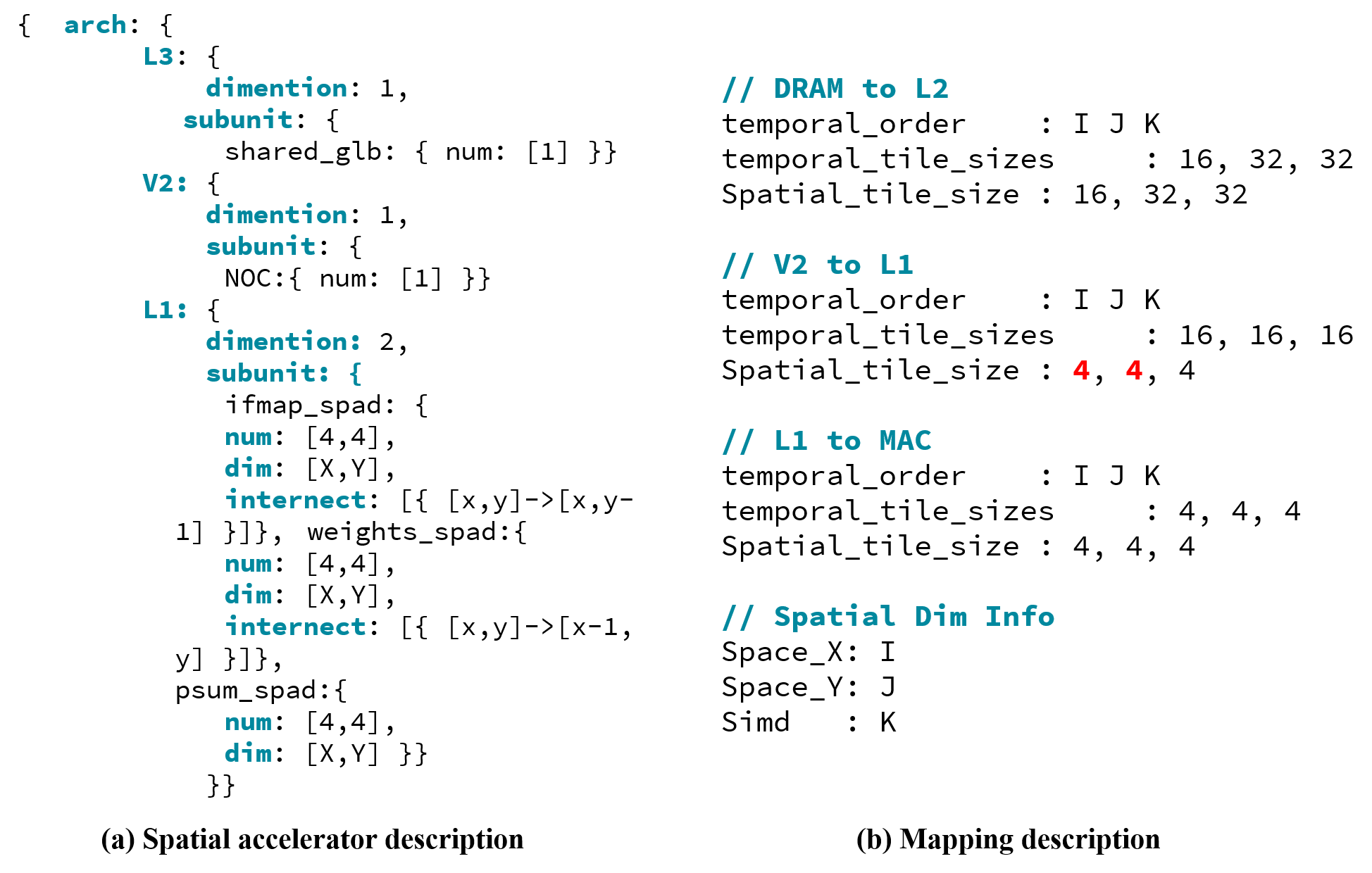}}
	\caption{Example of the spatial architecture description}
	\label{fig:description}
\end{figure}

PolyAcc uses memory-centric nations to describe the spatial accelerator, which is a hierarchical tree to abstraction like the previous work\cite{timeloop}.
Fig.\ref{fig:description}(a) shows the abstract of the example spatial accelerator in Fig.\ref{fig:example}(a) using tree-like description.
We introduce three attributes to represent the topology of interconnection.
The first attribute, Dim indicates how the current memory level units are laid in the physical dimension, by default on the x-axis. 
The second attribute, Virtual, facilitates the description of the spatial mapping of data tile, and the memory level with this attribute has no dedicated physical memory. 
The third attributes, Connect, defines the interconnection of units at the current memory level, we use the Presburger Relations to describe the connection as below: 
\begin{equation}\label{connection}
connection = \{[\overrightarrow{s_1}]\to[\overrightarrow{s_2}]\} 
\end{equation}

where $\overrightarrow{s_1}$ and $\overrightarrow{s_2}$ denote the coordinate of the hardware units that are connected. 
The relation can represent connections within the same level of hardware or connections at different levels of hardware. As shown in Fig.\ref{fig:description}(a), the ifmap\_spad in  L1 level memory has connect relation attribute:$\{[x,y]\to [x,y-1]\}$, which means that data in the ifmap\_spad can be propagated down to the adjacent ifmap\_spad.

% (b)Abstraction for Mapping
% The data partitioning and scheduling strategies used by spatial accelerators are known as dataflow, we call mapping in this artical. 
We target the data tile to describe the spatial/temporal tile relations use loop-centric approach like Union\cite{union}. In the mapping abstraction, the meanings of the description notations we used are summarized in Table~\ref{tab:notation}. 
% temporal\_ordal defines the temporal ordring between in the memory level. 
% tempral\_tile\_size and spatial\_tile\_size define the size of temporal and spatial tile for each dimension.

\newcommand{\tabincell}[2]{\begin{tabular}{@{}#1@{}}#2\end{tabular}}
\begin{table}[h]
    \centering
    \footnotesize
    \renewcommand{\arraystretch}{1.5}
    \setlength{\tabcolsep}{3pt}
    \caption{Description of notations used in abstraction for Mapping.}
    \label{tab:notation}
    \scalebox{.8}{
    \begin{tabular}{c|l}
    \hline
    Nation & \multicolumn{1}{c}{Description} \\ \hline
    temporal\_ordal     & \tabincell{c}{the temporal ordering between in current memory level}                            \\
    tempral\_tile\_size & \tabincell{c}{the size of temporal tile  for each dimension \\ in current memory level}             \\
    spatial\_tile\_size & \tabincell{c}{the size of temporal and spatial tile for each dimension \\in current memory level} \\
    Space\_X, Space\_Y  & \tabincell{c}{the parallelizable dimension map to the physical x and y axes}                    \\
    SIMD   & vectorized dimension \\ \hline
    \end{tabular}
    }
\end{table}
To avoid ambiguity, we use Space\_X and Space\_Y to describe the parallelizable dimension map to the physical x and y axes.
% And SIMD attribute to describe which dimension is SIMD vectorized.
After determining the space loop, we need to determine the parallelism of the space loop.
We define the \emph{k} dimension of spatial and temporal tile size in \emph{i}th level memory as $S^{k}_{i}$, $T^{k}_{i}$, which can be divide into smaller spatial tile size and distributed into multi instances of the level-(i-1) memory. 
if $\frac{T^{k}_{i}}{S^{k}_{i}}$ is greater than one, it means that the \emph{\textbf{k}} dimension in the level-i memory is parallel, and the parallelism is $\frac{T^{k}_{i}}{S^{k}_{i}}$. A example mapping for MM operator is shown in Fig.\ref{fig:description}(b), we notice that $\frac{T^{I}_{L1}}{S^{I}_{L1}}=4$ and $\frac{T^{J}_{L1}}{S^{J}_{L1}}=4$, which means I, J are parallel, and the parallelism of the I, J dimensions is 4,4, respectively.

% I, J are respectively mapped to the physical X-axis and Y-axis in the next memory level contacting the \emph{Space} attribute 

Before evaluating the mapping performance, we need to check the legality of the mapping with several rules: 
(1) the execution order of statement instances must not violate the original statement dependencies. 
(2)the parallelism for the \textbf{k} dimension at i-level memory, which should be equal to or small then the number of (i-1)-level memory. 
(3)the tile size must meet the hardware specifications and be less than or equal to the capacity of the corresponding memory.

We integrate the mapping description with Schedule Trees\cite{schedule_tree} to represent the lexicographical order of the workload. The schedule tree starts with a domain node that defines the iteration domain of the program, followed with band nodes that encodes the partial schedule at each loop dimension\cite{autosa}.

\begin{figure}[h]
	\centerline{\includegraphics[width=\linewidth]{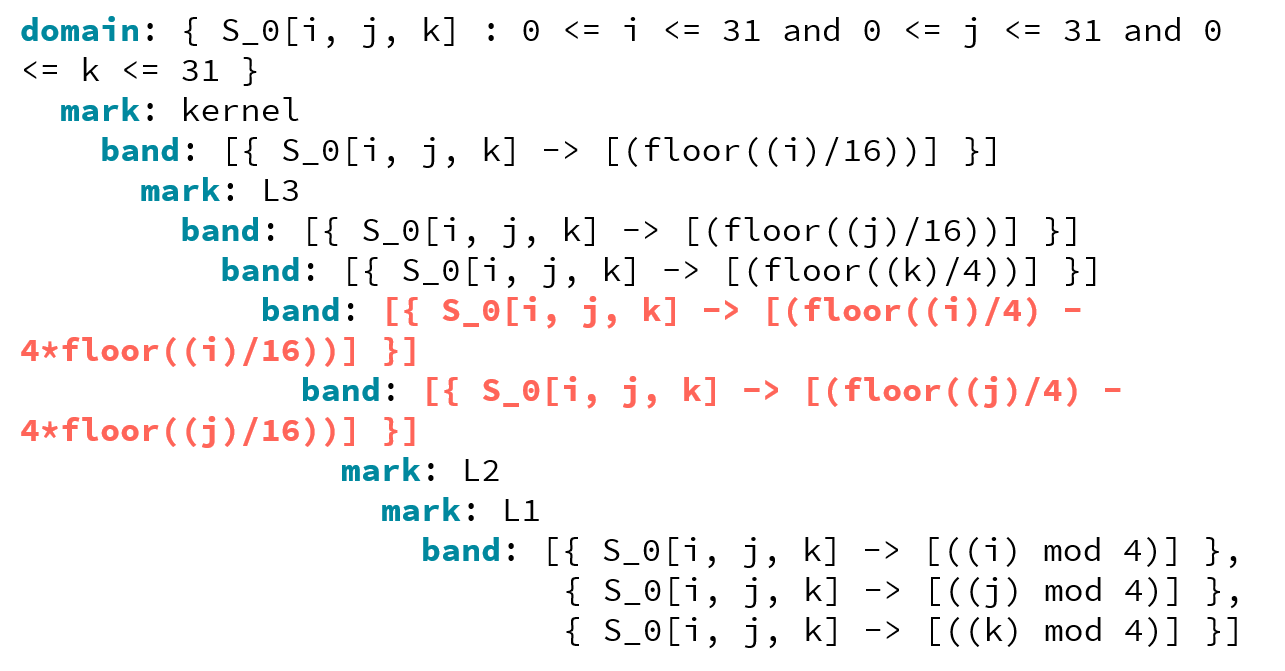}}
	\caption{Schedule tree of MM after loop transformation}
	\label{fig:sch_tree_after_trans}
\end{figure}

% We manipulates the schedule tree using the Integer Set Library, e.g., loop band tile, loop interchange. 
% We apply the MM mapping abstraction in Fig.\ref{fig:Mapping}(a) to the initial scheduling tree in Fig.\ref{fig:Mapping}(c) and get the final scheduling tree in Fig.\ref{fig:sch_tree_after_trans}.

We construct a schedule tree that includes the memory hierarchy using mapping, as shown in Fig.\ref{fig:sch_tree_after_trans}.
We mark the memory level using \textbf{mark} node, indicating that the subtree band is associated with the next level of memory. 
% A subtree marked by a \textbf{mark} node is then isolated from its original position in the schedule tree, thereby obtaining the temporal and spatial affine of each level of memory. 
In addition, we add an attribute of the scheduling tree band to indicate whether the loop is a time loop or a space loop. Due to space limitations, we do not mark it in Fig.\ref{fig:sch_tree_after_trans} but marked the space loop band in red color, and the time loop band in black color.

\subsection{Data Placement Relation}
The major feature of our polyAcc is to capture the runtime behavior of all the physical units in the hardware by data placement relation.
Our data placement relation applies an affine function to capture the space-time hierarchy, each $(\overrightarrow{s}, \overrightarrow{t})$ pair corresponds to a memory level in the hardware, where  $\overrightarrow{s}$ and $\overrightarrow{t}$ are multi-dimensional vector to describe space-stamp and time-stamp. 
The $\overrightarrow{s}$ describes hardware unit coordinates where statement instances are executed. 
The $\overrightarrow{t}$ determines the execution sequence, a classical lexicographical ordering to realize relative ordering among timestamps when statement instances are executed in the current memory level. 

We define \emph{space-time map} of the level-i memory as follows:
\begin{equation}\label{st}
% \nonumber
ST = \{S[\overrightarrow{n}]\to[[ \overrightarrow{s} ] \to [\overrightarrow{t}]]\}
\end{equation}
% where $\overrightarrow{s}$, $\overrightarrow{t}$ represent space-stamp and time-stamp vector of the hardware unit, and $S[\overrightarrow{n}]$ represent the loop instance. 
The equation~\ref{st} denotes that loop instance or loop tile $S[\overrightarrow{n}]$ is accessed by hardware unit $\overrightarrow{s}$ in the level-i memory at the time-stamp  $\overrightarrow{t}$.

% \begin{figure}[ht]
%     \centering
%     \includegraphics[width=\linewidth]{mapping_st.png}
%     \caption{(a) the space-time map for each memory level for example architecture and mapping. (b)the example of the data partitioning and scheduling strategy for example mapping.}
%     \label{fig:dpa}
% \end{figure}

We scan the schedule tree from top to bottom, using the time band as the time-stamp vector $\overrightarrow{t}$ and the space band as the space-stamp vector $\overrightarrow{s}$ for the memory level represented by the mark node;
if there is no space band, the space-stamp vector is constant 1.
% As shown in Fig.\ref{fig:dpa}(b), we get the space-time map of each level of memory by scanning the scheduling tree in Fig.\ref{fig:sch_tree_after_trans} from top to bottom. 
% Note that there is no time band between L1 and L2 nodes in the scheduling tree, the time-stamp L1 is $\lfloor i/4\rfloor  - 4*\lfloor i/16\rfloor + \lfloor j/4\rfloor + \lfloor k/4\rfloor  - 3*\lfloor i/16\rfloor$ instead of empty because of the interconnection between units at the L1 level.
When calculating the affine expression of the time-stamps, the lexicographic order of the loop instances and the inter-unit's connections in the current memory level need to be considered.

We can combine the Access Function \textbf{R} and space-time map \textbf{ST} to model the array reference mapping in time-stamp and space-stamp at level-i memory. 
We define the data placement relation $\theta$ of array $M$ of level-i memory:
\begin{equation}\label{dpa_m}
\begin{split}
% \nonumber
\theta^M_i = \{M[\overrightarrow{p}] \to [[\overrightarrow{s_i} ] \to [\overrightarrow{t_i}]] \}
\end{split}
\end{equation}
where $\overrightarrow{s_i}$, $\overrightarrow{t_i}$ represent the space-stamps and time-stamps of the i-level memory hardware unit. Equation~\ref{dpa_m} indicates that array elements $M[\overrightarrow{p}]$ appear in i-level memory hardware unit $[\overrightarrow{s_i} ]$ at time $[\overrightarrow{t_i}]$. 

% The data partitioning and scheduling strategy in Fig.\ref{fig:description}(b) is shown in Fig.\ref{fig:dpa}(b). The DRAM is connected to the global shared buffer (L3 level memory), data can move from the DRAM to the global shared buffer. 
% Moreover, the data in the global shared buffer can be multicast to the register files in different PEs through the on-chip network (Virtual L2 memory), and the MAC in the PE can access the data in the register files (L1 level memory).
Considering the impact of memory in the accelerator, We define the data placement relation $\Theta$ to capture the moving relation from (i+1)-level memory to i-level memory:
\begin{equation}\label{DPA}
% \nonumber
\Theta^M_i = \{M[\overrightarrow{p}] \to [[[\overrightarrow{s_{i+1}}]\to[\overrightarrow{t_{i+1}}]] \to [[\overrightarrow{s_{i}}]\to[\overrightarrow{t_{i}}]]\}
\end{equation}
where $s_{i+1}$, $t_{i+1}$ and $s_{i}$, $t_{i}$ represent space-stamp and time-stamp of the parent level and child level.

The data placement relation $\theta$  provides a relation to data on how to place in space-stamps and time-stamps. For a given mapping, the space-time map $ST$ and the array access function of the workload together can precisely identify what array elements set must appear at each space-time coordinate of each hardware unit.

\subsection{Performance model}
Based on the data placement relation, we know that when the data is placed in which unit in the memory level. 
We can precisely compute various performance metrics using the ISL and Barvinok library. Below, we introduce the performance metrics involved in this article, including data reuse, latency, and energy.

\subsubsection{Data Reuse and Unique Volume}
The memory hierarchies and dataflow of spatial accelerators lead to different data access patterns. Evaluating the mapping for metrics starts with defining several data volumes, including different data reuse volume and unique volume.

\textbf{1)Data Reuse Volume} 
is the amount of data reuse across multiple time-stamps or space-stamps, including temporal reuse volume~(TV), spatial reuse volume~(SV) and temporal-spatial reuse volume~(TSV).

Algorithm~\ref{alg:data_reuse} describes the detailed procedure of calculating the data volumes for different access pattern, We derive some R-transfer relations(line~\ref{alg:line:t_r} - line~\ref{alg:line:ts_r}) to exploit data reuse at the current level. R-transfer relations specify a schedule for moving data instance or data tiles from parent unit to child unit.
Once R-transfer relations are determined, we derive activity counts representing the workload’s execution
(eg. TV, SV, TSV).

We use temporal reuse as a example to illustrated how to compute data reuse volume. 
% We derive a R-transfer relations(line~\ref{alg:line:t_r}) that takes into account the fact that data that was
% already present in the buffer at time need not be fetched
% from anywhere at time + 1.
The R-transfer relations(line~\ref{alg:line:t_r}) about temporal reuse in the level-i memory as follow: 
\begin{equation}\label{eq:r_t}
\begin{split}
% \nonumber
R\_t = \{[[[\overrightarrow{s_{i+1}}]\to[\overrightarrow{t_{i+1}}]]\to[[\overrightarrow{s_{i}}]\to[\overrightarrow{t_{i}}]]\to\\
[[[\overrightarrow{s_{i+1}}]\to[\overrightarrow{t_{i+1}}]]\to[[\overrightarrow{s_{i}}]\to[\overrightarrow{t_{i}}]]]:\overrightarrow{t_{i}}\ll\overrightarrow{t\prime_{i}}\}
\end{split}
\end{equation}
where $\overrightarrow{s_{i+1}}$, $\overrightarrow{t_{i+1}}$ and $\overrightarrow{s_{i}}$, $\overrightarrow{t_{i}}$ reprent the space-stamp and time-stamp of the (i+1)-level and i-level memory respectively. The operator $\ll$ represents the closest time vector in their lexicographic orderings.

The level-(i+1) is parent level of and level-i memory, which means the level-i unit $\overrightarrow{s_{i}}$ fetch data from $\overrightarrow{s_{i+1}}$ in time $\overrightarrow{t_{i}}$. $R\_t$ takes into account the fact that date was already present in the $\overrightarrow{s_{i}}$ at time $\overrightarrow{t_{i}}$ need not be fetched from anywhere at time $\overrightarrow{t\prime_{i}}$. 
We have similar R-transfer relations for other forms of reuse such as spatial reuse and temporal-spatial reuse.

\textbf{2) Data Unique Volume} is the number of unique data at the current memory level. Data is considered unique if it can not fetched from adjacent time-stamps at the same space-stamps. It can be calculated by the difference in DPR of adjacent time-stamps. 

\begin{algorithm}[t]
\caption{Calculate the data volumes for different access pattern at $level-i$ memory}
\label{alg:data_reuse}
    \begin{algorithmic}[1]
        \REQUIRE The DPA of array M in the level-i memory, $\Theta^M_{i}$;\\
        % The inter-connection relation of $level-i$ memory, $C_i$
        % The array accesses function $F$
		\ENSURE The number of temporal reuse volume, $TV$;\\
		The number of spatial reuse volume, $SV$;\\
		The number of temporal-spatial reuse volume, $TSV$;\\
		The number of unique volume, $UV$;\\
		The number of total memory access count, $Total$;
		\STATE{the DPA is \\$\Theta^M_i = \{M[\overrightarrow{p}] \to [[[\overrightarrow{s_{i+1}}]\to[\overrightarrow{t_{i+1}}]]\to[[\overrightarrow{s_{i}}]\to[\overrightarrow{t_{i}}]]\}$}
		\label{alg:dpa}
		\STATE{$R\_t = \{[[[\overrightarrow{s_{i+1}}]\to[\overrightarrow{t_{i+1}}]]\to[[\overrightarrow{s_{i}}]\to[\overrightarrow{t_{i}}]]
        \to[[[\overrightarrow{s_{i+1}}]\to[\overrightarrow{t_{i+1}}]]\to[[\overrightarrow{s_{i}}]\to[\overrightarrow{t_{i}}]]]:\overrightarrow{t_{i}}\ll\overrightarrow{t\prime_{i}}\}$}
        \label{alg:line:t_r}
		\STATE {$R\_s = \{[[[\overrightarrow{s_{i+1}}]\to[\overrightarrow{t_{i+1}}]]\to[[\overrightarrow{s_{i}}]\to[\overrightarrow{t_{i}}]]\to[[[\overrightarrow{s_{i+1}}]
		\to[\overrightarrow{t_{i+1}}]]\to[[\overrightarrow{s\prime_{i}}]\to[\overrightarrow{t_{i}}]]]:\overrightarrow{s_{i+1}}\ll\overrightarrow{s\prime_{i}}\land\overrightarrow{s_{i}}\gg\overrightarrow{s\prime_{i}}\}$}
		\label{alg:line:s_r}
		\STATE {$R\_st = \{[[[\overrightarrow{s_{i+1}}]\to[\overrightarrow{t_{i+1}}]]\to[[\overrightarrow{s_{i}}]\to
		[\overrightarrow{t_{i}}]]\to[[[\overrightarrow{s_{i+1}}]\to
		[\overrightarrow{t_{i+1}}]]\to[[\overrightarrow{s\prime_{i}}]\to[\overrightarrow{t\prime_{i}}]]]:\overrightarrow{t_{i}}\ll \overrightarrow{t\prime_{i+1}} \land ((\overrightarrow{s_{i}} \ll \overrightarrow{s\prime_{i}}) \lor (\overrightarrow{s_{i}} \gg \overrightarrow{s\prime_{i}}))\}$}
		\label{alg:line:ts_r}
		\STATE {${\Delta_t} = \Theta^M_i \cap (\Theta^M_i \circ R\_t)$}
		\STATE {${\Delta_s} = \Theta^M_i \circ R\_t$}
		\STATE {${\Delta_{st}} = \Theta^M_i \cap (\Theta^M_i \circ R\_st)$}
		\STATE {$Total = sum(\Theta^M_i)$}
		\STATE {$TV = sum(\Delta_t)$}
		\STATE {$SV = sum(\Delta_s)$}
		\STATE {$TSV = sum(\Delta_{st})$}
		\STATE {$UV = Total-TSV$}
		\STATE \textbf{return} {$TV$, $SV$, $TSV$, $UV$, $Total$}
\end{algorithmic}
\end{algorithm}

\subsubsection{Utilization of PE Array}
PE array utilization can be computed by dividing the number of active MAC units by the total number of physical MAC units, utilization is only related to spatially unrolled loops. 
\begin{equation}\label{eq:Util}
Util = \frac{Total_{mac}}{PE\_size}
\end{equation}

\subsubsection{Estimating Energy Consumption}
The energy consumption is an essential metric for accelerator. We take into account MAC computation energy and memory access energy. For each operator, the total energy is the sum of activate and idle components of MAC, on-chip scratchpad access energy, main memory DRAM access energy and interconnect energy, expressed as follows:
\begin{equation}\label{eq:eaergy}
E_{total} = E_{mac} + E_{on-chip} + E_{DRAM} + E_{connect} 
\end{equation}

\textbf{1) Compute Energy.} We define MAC computation energy as computing energy consisting of two components: the activate MAC energy and the idle MAC energy. 
The active energy consumed in performing a MAC operation and idle energy consumed per compute unit when not actively performing a compute operation.
Total compute energy is calculated as follows:
\begin{equation}\label{qe:mac_energy}
\begin{split}
E_{mac} = (Util \times e_{act} + (1 - Util) \times e_{idle})\times Total_{mac}  
\end{split}
\end{equation}
where $Util_{PE}$ is utilization of PE array, 
the $e_{activate}$ is active energy consumed by the MAC per operation, and $e_{idle}$ is energy consumed when MAC is inactive.

\textbf{2) On-chip scratchpad Energy.} 
On-chip scratchpad energy consists of read energy and write energy as follows:
\begin{equation}\label{eq:on_chip_energy}
E_{On-chip} = \sum_{op=1}^{operands}{e_{w} \times UV^{op} + e_{r}\times (UV^{op}_{child} + TV^{op}_{child})}
\end{equation}
where $e_{w}$, $e_{r}$ are the energy consumed by each write operation and read operation of the on-chip scratchpad, $y$ represent array index. 
Considering the impact of data reuse, 
the write energy is the product of $e_{w}$ and the unique volume (UV); the read energy is the product of $e_{r}$ and the unique volume (UV), temporal reuse volume (TV) of the child-level memory.

\textbf{3) Main Memory Energy.} 
We derive the energy consumption of main memory as follows:
\begin{equation}\label{eq:dram_energy}
E_{dram} = \sum_{op=1}^{operands}{e_{w} \times UV^{op} + e_{r}\times (UV^{op} + TV^{op})}
\end{equation}
where $e_{w}$, $e_{r}$ are the energy consumed by each write operation and read operation of the DRAM, $y$ represent array index. UV and TV are the unique volume, temporal reuse volume of memory connected to DRAM.

\textbf{4) Connect Energy.} 
We model the energy consumed by connects that multi-cast connects and inter-connect for each level memory as follows:
\begin{equation}\label{eq:interconnect_energy}
E_{connect} = \sum_{op=1}^{operands}{(e_{multi} \times SV^{op}_i + e_{inter} +TSV^{op}_i)}
\end{equation}
Where $e_{multi}$ and $e_{inter}$ are the energy consumed by multi-cast connects and inter-connect from parent-level memory to child-level memory. 
The total connect energy consumption is the sum of the per-data-access energy multiplied by the spatial reuse volume~$SV$ and the temporal-spatial reuse volume~$TSV$ for each memory level.

\subsubsection{Execution Time}
The execution time of spatial accelerator consists of communication and computation time.
We assume accelerator works in a pipeline fashion, the overall execution time is just the maximum of communication time $cycles\_comm$ and computation time $cycles\_comp$.
\begin{equation}\label{eq:total_cycle}
total\_cycles = max(cycles\_comp, cycles\_comm)
\end{equation}

\textbf{1) computation time.} 
The computation time can be estimated by the total number of mac operations and the average number of active PEs, it is calculated as follows:
\begin{equation}\label{eq:compute_cycle}
cycles\_comp = \frac{total_{mac}}{act_{PE}} \times lat_{avg}
\end{equation}
where $active_{PE}$ is the number of activate PEs across time-stamps, and $lat_{avg}$ is average time of MAC operation.

\textbf{2) communication time.} 
To model communication time, we takes into account the fact that the data are fetch from DRAM to on-chip scratchpad, then the data in on-chip scratchpad are broadcast to private register files (RF) in PE.
The communication time consists of two parts, the time for DMA access to DRAM and the time for data transfer in the on-chip scratchpad.

First, we calculate the time for DMA access to DRAM. The volume of data transferred from the DRAM to the on-chip scratchpad is estimated by the UV of operands; 
in other words, the total burst size of DMA is UV.
% Total DMA invocations can be obtained by time elapse and reuse factor.
To calculate DMA cycles, we consider a latency model of Cell processors~\cite{Cell}, i.e.,
\begin{equation}\label{eq:DMA}
\begin{split}
% \begin{aligned}
cycles_{DMA}^{op} = (reqs \times init + 0.25 \times UV_{op}) \times \frac{f_{accel}}{f_{dma}}
\end{split}
\end{equation}
where $reqs$ is the total DMA invocations required, $init$ is DMA latency, $f_{accel}$ and $f_{dma}$ are the frequencies of the DMA and accelerator respectively. And, we calculate the cycles required for accessing the DRAM as follows:
\begin{equation}\label{eq:time_dram}
\begin{split}
cycles_{DRAM} = \sum_{op=1}^{operands}{cycles_{DMA}^{op}}
\end{split}
\end{equation}

Second, we calculate the time for data transfer between the on-chip scratchpad and RFs in PEs. 
While PEs process data from RFs, if operands are not reused, new data for the next new data tile needs to be fetched from the on-chip scratchpad and communicated to PEs via interconnect. 
The time for data transfer via the interconnect of PEs and the computation performed by the PE are coincident. 
Therefore, we consider the time of multicasting or unicasting the data from the on-chip scratchpad to the RF. 
The cycles required for multicast and unicast operands are respectively:
\begin{equation}\label{eq:spatial_comm}
\begin{split}
cycles_{multicast}^{op} = \frac{SV}{B}
\end{split}
\end{equation}

\begin{equation}\label{eq:unique_comm}
\begin{split}
cycles_{unicast}^{op} = \frac{max(Total-SV-TV,0)}{B}
\end{split}
\end{equation}
where $B$ is the width of the data bus for interconnect. Total cycles required to transfer data between the on-chip scratchpad and RFs is
\begin{equation}\label{eq:on_chip_comm}
\begin{split}
cycle_{on-chip} = \sum_{op=1}^{operands}{cycles_{multicast}^{op} + cycles_{unicast}^{op}}
\end{split}
\end{equation}

Based on the cycles required for accessing the DRAM and the time for data transfer from the scratchpad to RFs, we calculate the total cycles required for communication as follows:
\begin{equation}\label{eq:time_dram}
\begin{split}
cycles\_comm = max(cycles_{DRAM}, cycle_{on-chip})
\end{split}
\end{equation}

\section{EVALUATION}
\subsection{Experiment Setup And Analysis}
\textbf{Benchmarks} GEMM and 2D-Convlution operator are widely used in deep learning, scientific computations, We evaluate PolyAcc with them, Table~\ref{tab:Benchmark} list 6 dataflows that we use to evaluate our framework, including corresponding workloads and PE sizes. 
The dataflow is mainly named according to the spatial loop executed in the space, such as the \textbf{(IJ-sp)} data-flow in the GEMM operator, which means that PEs are grouped based on the unrolled I and J loops for spatial execution. 

\begin{table}[h]
    \centering
    \footnotesize
    \renewcommand{\arraystretch}{1.5}
    \setlength{\tabcolsep}{3pt}
    \caption{Operator, Dataflow and PE size used in evaluation}
    \label{tab:Benchmark}
    \scalebox{.8}{
    \begin{tabular}{|c|c|c|c|}
    \hline
    Benchmark & Problem size  & dataflow(Sp:Space loop) & PE Size   \\ \hline
    \multirow{3}{*}{GEMM}    & \multirow{3}{*}{[i,j,k]:[256,256,256]} & Output-Satation(IJ-Sp) & $8\times8$   \\ \cline{3-4} 
              &               & Weight-Station(KJ-Sp)   & $8\times8$  \\ \cline{3-4} 
              &               & Vector(J-Sp)            & $1\times64$ \\ \hline
    \multirow{3}{*}{2D-CONV} & AlexNet-conv2\cite{Alexnet}                          & ROW-Station($O_{y}O_{y}-Sp$)\cite{eyeriss} & $14\times12$ \\ \cline{2-4} 
              & MobileNetV2-2\cite{MobileNetV2} & Weight-Satation(KC-Sp)  & $8\times8$  \\ \cline{2-4} 
              & ResNet50-1\cite{Resnet50}    & Shi-diannao($O_{x}O_{y}-Sp$)\cite{ShiDianNao}  & $8\times8$  \\ \hline
    \end{tabular}
    }
\end{table}

\begin{figure}[ht]
    \centering
    \includegraphics[width=\linewidth]{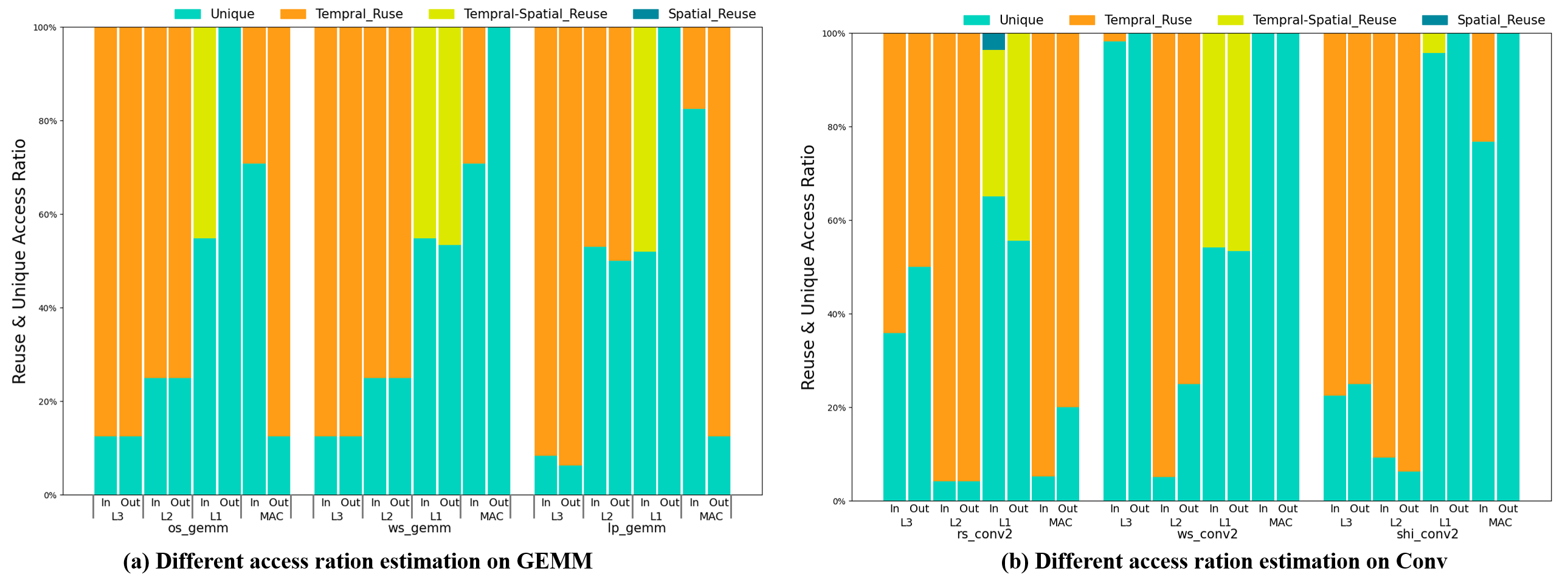}
    \caption{The result of data access pattern in different dataflow for GEMM and 2D-Conv respectively.}
    \label{fig:access_ration}
\end{figure}

\begin{figure}[ht]
    \centering
    \includegraphics[width=\linewidth]{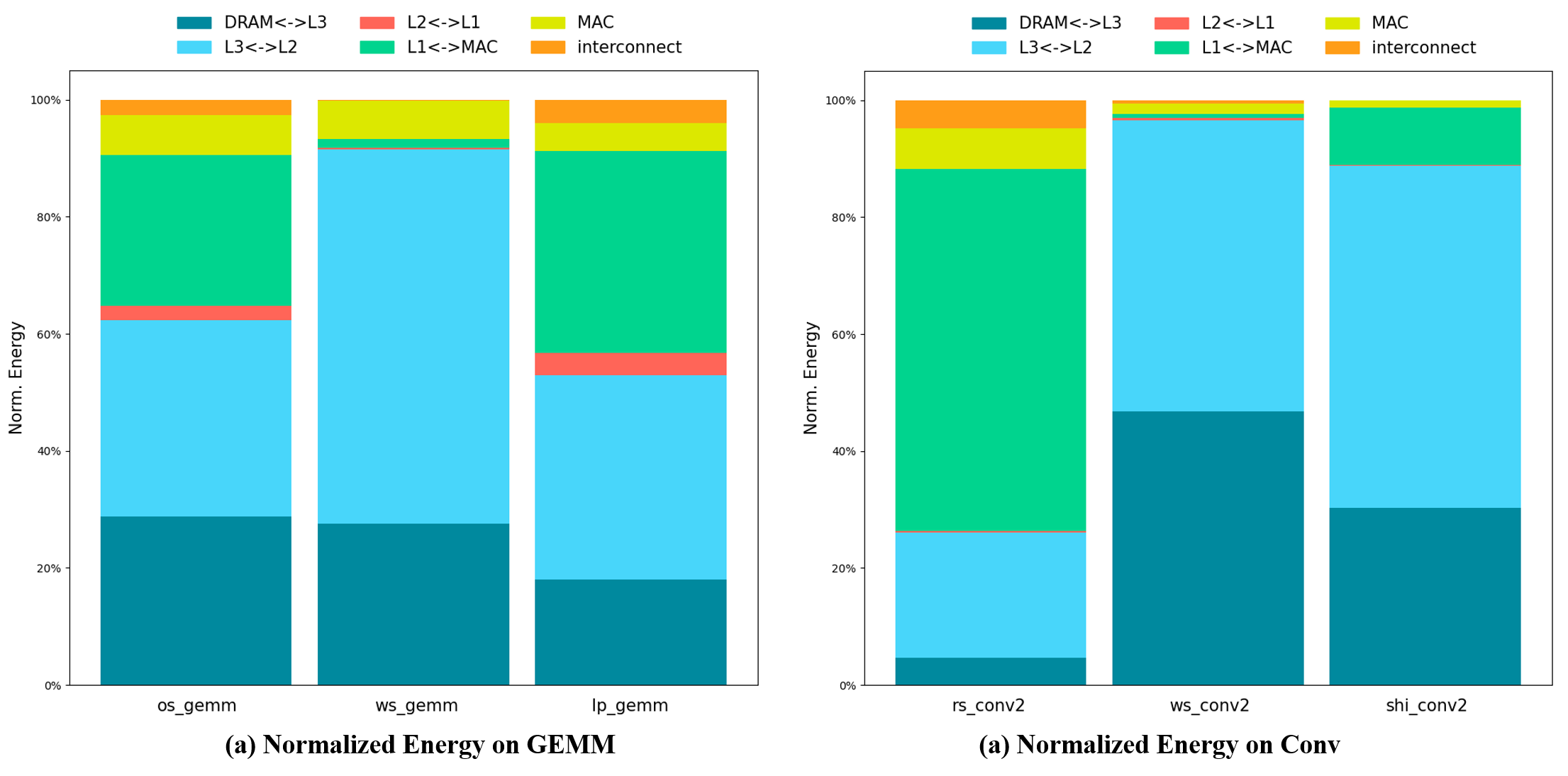}
    \caption{Energy analysis for GEMM and 2D-Conv, respectively.}
    \label{fig:energy}
\end{figure}

Fig.\ref{fig:access_ration} shows the data access pattern in different level memory for GEMM and 2D-Conv. We find that different dataflows have high variation in data access patterns because of the interconnect of PEs in different dataflows. We observe that the data access pattern of each memory level is different. For example, The data access pattern at the L3 level is mainly the temporal reuse since  L3-level memory as a data buffer fetches data from DRAM, reducing DRAM access and energy consumption. 

Moreover, Fig.\ref{fig:energy} shows the breakdown of the energy for system resources. It noted that the different colors in the bars represent the energy consumption of different memory levels or interconnects. We observe that the energy consumption is attributed to accessing data from the private/shared scratchpad, and the unique access of the child memory level affects the energy consumption of data access at the parent memory level.
For optimized the execution of workloads on the spatial accelerator, the dedicated memory resources should be fully utilized for data reuse as much as possible, reducing the energy consumption of unique accessing.

\begin{figure}[ht]
    \centering
    \includegraphics[width=0.9\linewidth]{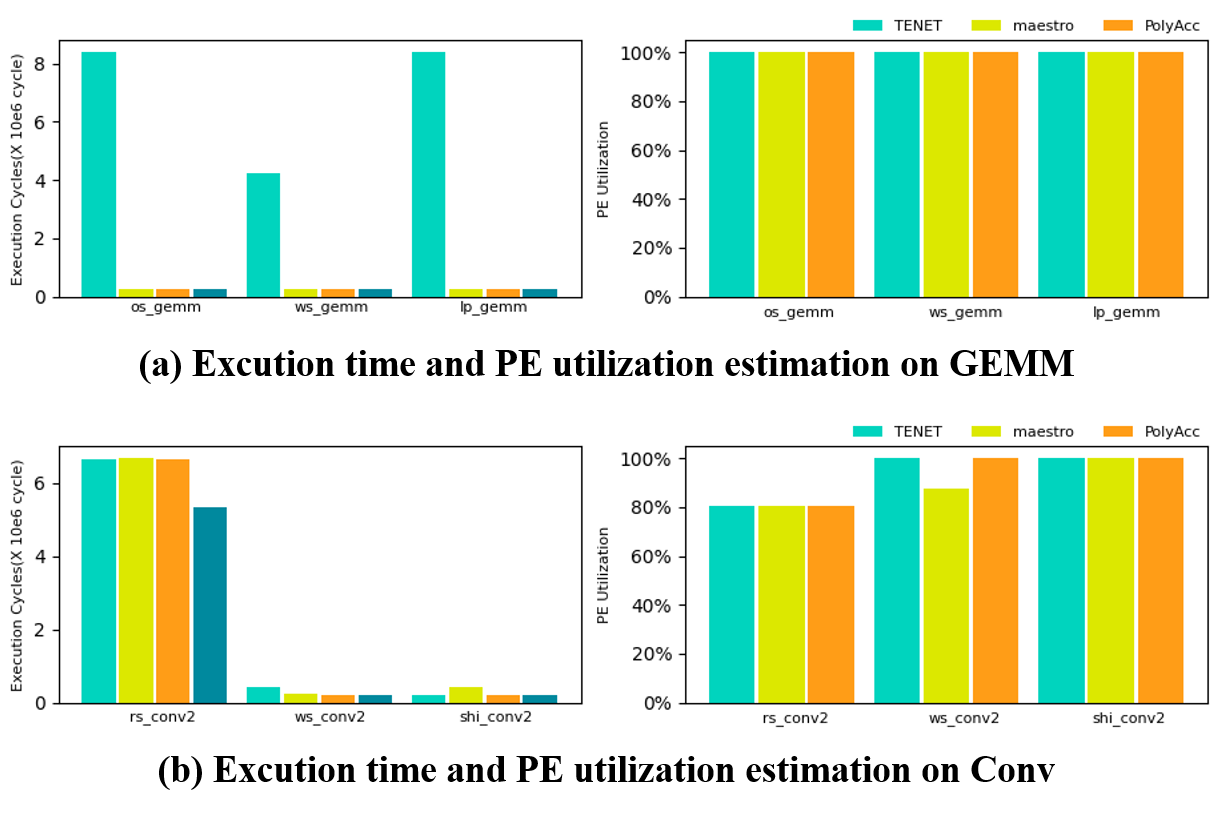}
    \caption{Execution time and PE utilization comparison with MAESTRO and TENET. (a) and (b) are the result of excution time and PE utilization for GEMM and 2D-Conv, respectively. (Right: Execution time; Left: PE utilization)}
    \label{fig:time_util}
\end{figure}

Fig.\ref{fig:time_util} compares the execution time and PE utilization of TENET and MASTERO on our benchmarks. We approximate the ground-truth value of execution time for estimated execution time using $\frac{total MAC operations}{totals PEs}$. Furthermore, we find that our estimations closely matched the execution time of the ideal value. For estimated execution time, The estimated results of the three frameworks are similar, with a difference of 12.5\% in the largest.

\subsection{Estimation Comparison}
We comparison our framework with the recent work\cite{tenet, maestro}, which are the state-of-the-art relation-centric and data-centric notation with comprehensive performance analysis.
We show that our proposed model improves the metric estimation accuracy.

We considered a similar dataflow accelerator architecture as\cite{eyeriss}, which consists of $14\times12$ PEs with 16-bit precision, and it has two levels of physical memory(global buffer and register file inside the PE) and one level of virtual memory that model NOC, the multi-cast networks to communicate the operands to PEs.
We configure the them to use the same parameters (PE Number, Bandwidth, Buffer Size). For the interconnection parameters of PolyAcc, We define the connection relationship of the output feature downward, the weight to the left, and the input feature multicast communication along the diagonal line, respectively. 
% And, We use the eyeriss dataflow directives in maestro, since it maps workload at each logical cluster level under the assumption that each PE can communicate with neighboring PEs.

% This validation experiment covers various mapping mechanisms representing how operators are executed spatially by multi-dimensional space loop folding or unfolding to improve PE utilization. We used the exact mapping as \cite{eyeriss}, including the loop tile size and the spatial loop's choice. 
\begin{figure}[ht]
    \centering
    \includegraphics[width=\linewidth]{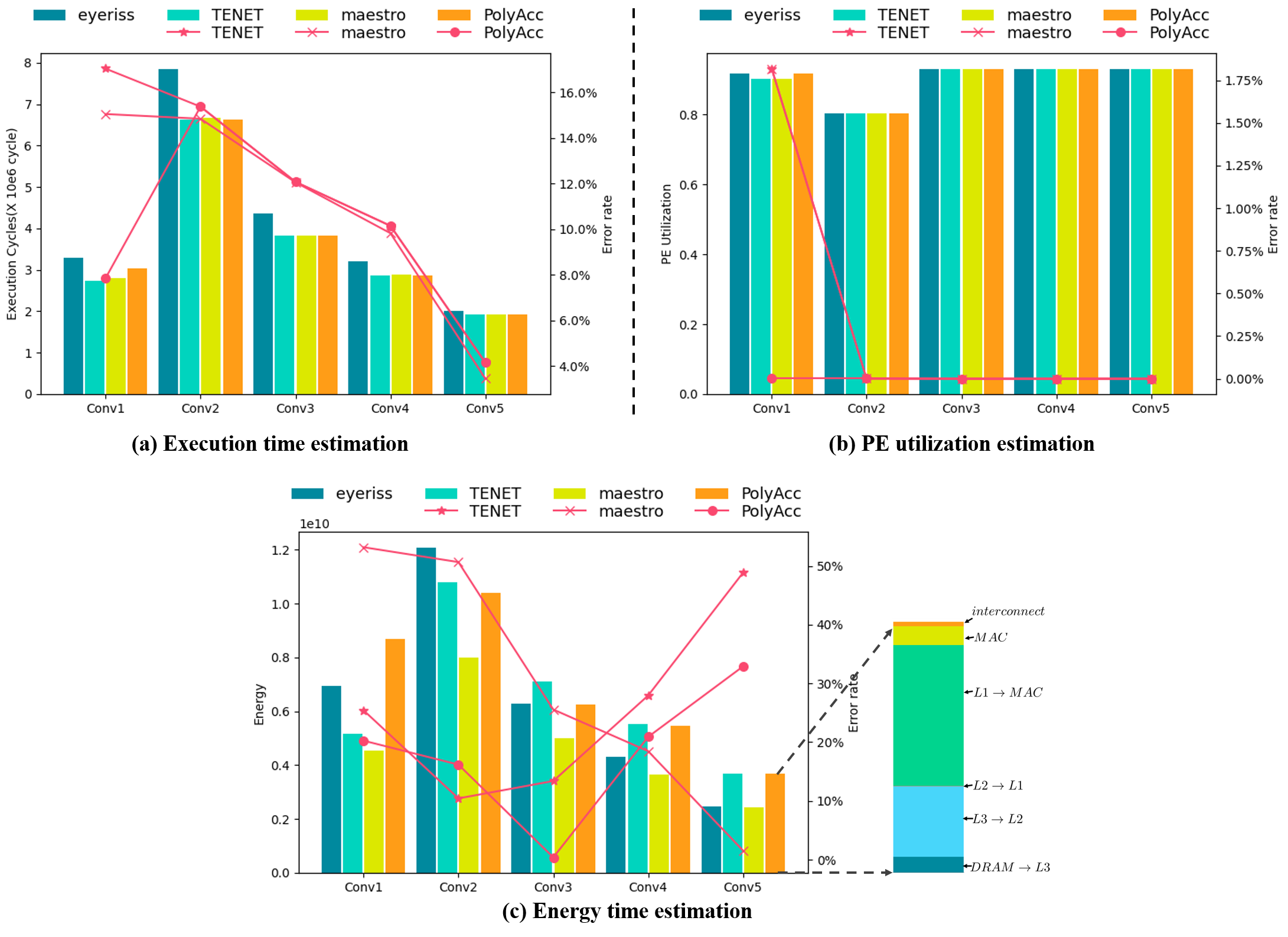}
    \caption{Latency, Energy, and PE utilization comparison with TENET and MAESTRO.(a), (b) and (c) are the result of Eyeriss-like\cite{eyeriss} accelerator running AlexNet\cite{Alexnet}, Conv1-Conv5 means convolutional layer in Alexnet.}
    \label{fig:Compared}
\end{figure}

We compare the execution time, PE utilization and energy calculated by our work and TENET, MAESTRO.
Fig.\ref{fig:Compared} shows the comparison results. 
PolyAcc achieves the same PE utilization as Yang as eyeriss and TENET. 
Moreover, for various mapping mechanisms, the execution time estimated by PolyAcc closely matched execution cycles of the accelerator~\cite{eyeriss} thanks to more accurate reuse analysis, the difference is from 4.15\% to 15.4\%. Furthermore, We entirely consider the energy cost of the different data access patterns in different memory levels, which supports calculate the energy of different memory level. The result shows that the energy estimated by PolyAcc closely matched the~\cite{eyeriss}, with a difference of 7.5\% in the energy consumption.

\section{Conclusion}
% For efficient execution of operators on spatial accelerators, it is crucial to evaluate the resource utilization, data reuse, latency, and energy, which can guide the search for the optimal operator mapping execution method. 
In this work, we propose PolyAcc evaluate the performance of workload mappings on spatial accelerators. We present analysis methods leverage data placement, which can accurately capture the runtime behavior of hardware units.
% We also present a relation for spatial accelerators, which fully consider the different level memory characteristics and the interconnection of units. 
We evaluate our framework on six benchmarks, and results show that polyAcc closely matched the ideal value's execution time and PE utilization.
On validation against Eyeriss architecture, PolyAcc achieves 0.82\% improvements in execution time and 18.8\% improvements in energy consumption estimates compared to the state-of-the-art framework thanks to sophisticated data reuse analysis.

% if have a single appendix:
%\appendix[Proof of the Zonklar Equations]
% or
%\appendix  % for no appendix heading
% do not use \section anymore after \appendix, only \section*
% is possibly needed

% use appendices with more than one appendix
% then use \section to start each appendix
% you must declare a \section before using any
% \subsection or using \label (\appendices by itself
% starts a section numbered zero.)
%

% \appendices
% \section{Proof of the First Zonklar Equation}
% Appendix one text goes here.

% % you can choose not to have a title for an appendix
% % if you want by leaving the argument blank
% \section{}
% Appendix two text goes here.

% use section* for acknowledgment
% \ifCLASSOPTIONcompsoc
%   % The Computer Society usually uses the plural form
%   \section*{Acknowledgments}
% \else
%   % regular IEEE prefers the singular form
%   \section*{Acknowledgment}
% \fi

% The authors would like to thank...

% Can use something like this to put references on a page
% by themselves when using endfloat and the captionsoff option.
\ifCLASSOPTIONcaptionsoff
  \newpage
\fi

% trigger a \newpage just before the given reference
% number - used to balance the columns on the last page
% adjust value as needed - may need to be readjusted if
% the document is modified later
%\IEEEtriggeratref{8}
% The "triggered" command can be changed if desired:
%\IEEEtriggercmd{\enlargethispage{-5in}}

% references section

% can use a bibliography generated by BibTeX as a .bbl file
% BibTeX documentation can be easily obtained at:
% http://mirror.ctan.org/biblio/bibtex/contrib/doc/
% The IEEEtran BibTeX style support page is at:
% http://www.michaelshell.org/tex/ieeetran/bibtex/
%\bibliographystyle{IEEEtran}
% argument is your BibTeX string definitions and bibliography database(s)
%\bibliography{IEEEabrv,../bib/paper}
%
% <OR> manually copy in the resultant .bbl file
% set second argument of \begin to the number of references
% (used to reserve space for the reference number labels box)
% \begin{thebibliography}{1}

% \bibitem{IEEEhowto:kopka}
% H.~Kopka and P.~W. Daly, \emph{A Guide to \LaTeX}, 3rd~ed.\hskip 1em plus
%   0.5em minus 0.4em\relax Harlow, England: Addison-Wesley, 1999.

% \end{thebibliography}
\bibliographystyle{IEEEtran}
\bibliography{IEEEabrv,refs}

% Generated by IEEEtran.bst, version: 1.14 (2015/08/26)
\begin{thebibliography}{10}
\providecommand{\url}[1]{#1}
\csname url@samestyle\endcsname
\providecommand{\newblock}{\relax}
\providecommand{\bibinfo}[2]{#2}
\providecommand{\BIBentrySTDinterwordspacing}{\spaceskip=0pt\relax}
\providecommand{\BIBentryALTinterwordstretchfactor}{4}
\providecommand{\BIBentryALTinterwordspacing}{\spaceskip=\fontdimen2\font plus
\BIBentryALTinterwordstretchfactor\fontdimen3\font minus
  \fontdimen4\font\relax}
\providecommand{\BIBforeignlanguage}[2]{{%
\expandafter\ifx\csname l@#1\endcsname\relax
\typeout{** WARNING: IEEEtran.bst: No hyphenation pattern has been}%
\typeout{** loaded for the language `#1'. Using the pattern for}%
\typeout{** the default language instead.}%
\else
\language=\csname l@#1\endcsname
\fi
#2}}
\providecommand{\BIBdecl}{\relax}
\BIBdecl

\bibitem{timeloop}
A.~Parashar, P.~Raina, Y.~S. Shao, Y.-H. Chen, V.~A. Ying, A.~Mukkara,
  R.~Venkatesan, B.~Khailany, S.~W. Keckler, and J.~Emer, ``Timeloop: A
  systematic approach to dnn accelerator evaluation,'' in \emph{2019 IEEE
  International Symposium on Performance Analysis of Systems and Software
  (ISPASS)}, 2019, pp. 304--315.

\bibitem{Interstellar}
X.~Yang, M.~Gao, Q.~Liu, J.~Setter, J.~Pu, A.~Nayak, S.~Bell, K.~Cao, H.~Ha,
  P.~Raina, C.~Kozyrakis, and M.~Horowitz, \emph{Interstellar: Using Halide's
  Scheduling Language to Analyze DNN Accelerators}, 2020, p. 369–383.

\bibitem{dMazeRunner}
S.~Dave, Y.~Kim, S.~Avancha, K.~Lee, and A.~Shrivastava, ``Dmazerunner:
  Executing perfectly nested loops on dataflow accelerators,'' \emph{ACM Trans.
  Embed. Comput. Syst.}, vol.~18, no.~5s, oct 2019.

\bibitem{maestro}
H.~Kwon, P.~Chatarasi, M.~Pellauer, A.~Parashar, V.~Sarkar, and T.~Krishna,
  ``Understanding reuse, performance, and hardware cost of dnn dataflow: A
  data-centric approach,'' ser. MICRO '52, 2019, p. 754–768.

\bibitem{tenet}
L.~Lu, N.~Guan, Y.~Wang, L.~Jia, Z.~Luo, J.~Yin, J.~Cong, and Y.~Liang,
  ``Tenet: A framework for modeling tensor dataflow based on relation-centric
  notation,'' in \emph{2021 ACM/IEEE 48th Annual International Symposium on
  Computer Architecture (ISCA)}, 2021, pp. 720--733.

\bibitem{schedule_tree}
S.~Verdoolaege, S.~Guelton, T.~Grosser, and A.~Cohen, ``Schedule trees,'' 01
  2014.

\bibitem{isl}
S.~Verdoolaege, ``isl: An integer set library for the polyhedral model,'' in
  \emph{Mathematical Software -- ICMS 2010}, K.~Fukuda, J.~v.~d. Hoeven,
  M.~Joswig, and N.~Takayama, Eds., 2010, pp. 299--302.

\bibitem{union}
G.~Jeong, G.~Kestor, P.~Chatarasi, A.~Parashar, P.-A. Tsai, S.~Rajamanickam,
  R.~Gioiosa, and T.~Krishna, ``Union: A unified hw-sw co-design ecosystem in
  mlir for evaluating tensor operations on spatial accelerators,'' in
  \emph{2021 30th International Conference on Parallel Architectures and
  Compilation Techniques (PACT)}, 2021, pp. 30--44.

\bibitem{autosa}
J.~Wang, L.~Guo, and J.~Cong, ``Autosa: A polyhedral compiler for
  high-performance systolic arrays on fpga,'' in \emph{The 2021 ACM/SIGDA
  International Symposium on Field-Programmable Gate Arrays}, ser. FPGA '21,
  2021, p. 93–104.

\bibitem{Cell}
M.~Kistler, M.~Perrone, and F.~Petrini, ``Cell multiprocessor communication
  network: Built for speed,'' \emph{IEEE Micro}, vol.~26, no.~3, pp. 10--23,
  2006.

\bibitem{Alexnet}
A.~Krizhevsky, I.~Sutskever, and G.~E. Hinton, ``Imagenet classification with
  deep convolutional neural networks,'' \emph{Commun. {ACM}}, vol.~60, no.~6,
  pp. 84--90, 2017.

\bibitem{eyeriss}
Y.-H. Chen, T.~Krishna, J.~Emer, and V.~Sze, ``Eyeriss: An energy-efficient
  reconfigurable accelerator for deep convolutional neural networks,'' in
  \emph{2016 IEEE International Solid-State Circuits Conference (ISSCC)}, 2016,
  pp. 262--263.

\bibitem{MobileNetV2}
M.~Sandler, A.~G. Howard, M.~Zhu, A.~Zhmoginov, and L.~Chen, ``Mobilenetv2:
  Inverted residuals and linear bottlenecks,'' in \emph{2018 {IEEE} Conference
  on Computer Vision and Pattern Recognition, {CVPR} 2018, Salt Lake City, UT,
  USA, June 18-22, 2018}.\hskip 1em plus 0.5em minus 0.4em\relax Computer
  Vision Foundation / {IEEE} Computer Society, 2018, pp. 4510--4520.

\bibitem{Resnet50}
K.~He, X.~Zhang, S.~Ren, and J.~Sun, ``Deep residual learning for image
  recognition,'' in \emph{2016 {IEEE} Conference on Computer Vision and Pattern
  Recognition, {CVPR} 2016, Las Vegas, NV, USA, June 27-30, 2016}.\hskip 1em
  plus 0.5em minus 0.4em\relax {IEEE} Computer Society, 2016, pp. 770--778.

\bibitem{ShiDianNao}
Z.~Du, R.~Fasthuber, T.~Chen, P.~Ienne, L.~Li, T.~Luo, X.~Feng, Y.~Chen, and
  O.~Temam, ``Shidiannao: shifting vision processing closer to the sensor,'' in
  \emph{Proceedings of the 42nd Annual International Symposium on Computer
  Architecture}, D.~T. Marr and D.~H. Albonesi, Eds., 2015, pp. 92--104.

\end{thebibliography}

% biography section
% 
% If you have an EPS/PDF photo (graphicx package needed) extra braces are
% needed around the contents of the optional argument to biography to prevent
% the LaTeX parser from getting confused when it sees the complicated
% \includegraphics command within an optional argument. (You could create
% your own custom macro containing the \includegraphics command to make things
% simpler here.)
%\begin{IEEEbiography}[{\includegraphics[width=1in,height=1.25in,clip,keepaspectratio]{mshell}}]{Michael Shell}
% or if you just want to reserve a space for a photo:

% \begin{IEEEbiography}{Michael Shell}
% Biography text here.
% \end{IEEEbiography}

% if you will not have a photo at all:
% \begin{IEEEbiographynophoto}{John Doe}
% Biography text here.
% \end{IEEEbiographynophoto}

% insert where needed to balance the two columns on the last page with
% biographies
%\newpage

% \begin{IEEEbiographynophoto}{Jane Doe}
% Biography text here.
% \end{IEEEbiographynophoto}

% You can push biographies down or up by placing
% a \vfill before or after them. The appropriate
% use of \vfill depends on what kind of text is
% on the last page and whether or not the columns
% are being equalized.

%\vfill

% Can be used to pull up biographies so that the bottom of the last one
% is flush with the other column.
%\enlargethispage{-5in}

% that's all folks
\end{document}